\documentclass[sigconf]{acmart}
\AtBeginDocument{%
  \providecommand\BibTeX{{%
    \normalfont B\kern-0.5em{\scshape i\kern-0.25em b}\kern-0.8em\TeX}}}

\setcopyright{none}
\usepackage{amsmath,amsfonts}
 \usepackage{array}
 \usepackage{stfloats}
\usepackage{url}
\usepackage{verbatim}
 \usepackage[noend]{algorithmic}
\usepackage[vlined,ruled,commentsnumbered,linesnumbered]{algorithm2e}
\usepackage{tabularx}
\usepackage{multirow}
\usepackage{sidecap}
\usepackage{subfig}
\usepackage{subfloat}
\usepackage{graphicx}
\usepackage{textcomp}
\usepackage{booktabs} 
\usepackage{pifont}
\usepackage{lipsum}
\usepackage[outline]{contour}
\usepackage{subfloat}
\usepackage{comment}
\usepackage{balance}

\usepackage{tikz}

\newcommand{\sys}{\textsc{AutoMarks}}%
\newcommand{\basegp}{ICMarks}
\newcommand{\baseconstcap}{Cell Scattering}%
\newcommand{\baseinvasivecap}{Buffer Insertion}%

\DeclareMathOperator*{\argmin}{arg\,min}
\AtBeginDocument{%
  \providecommand\BibTeX{{%
    Bib\TeX}}}

\begin{document}

\title{Automated Physical Design Watermarking Leveraging Graph Neural Networks}




\author{Ruisi Zhang}
\email{ruz032@ucsd.edu}
\affiliation{%
  \institution{UC San Diego}
  \streetaddress{9500 Gilman Drive}
  \city{La Jolla}
  \state{CA}
  \country{USA}
  \postcode{92093}
}

\author{Rachel Selina Rajarathnam}
\email{rachelselina.r@utexas.edu}
\affiliation{%
  \institution{UT Austin}
  \streetaddress{9500 Gilman Drive}
  \city{Austin}
  \state{TX}
  \country{USA}
  \postcode{92093}
}

\author{David Z. Pan}
\email{dpan@ece.utexas.edu}
\affiliation{%
  \institution{UT Austin}
  \streetaddress{9500 Gilman Drive}
  \city{Austin}
  \state{TX}
  \country{USA} 
  \postcode{92093}
}

\author{Farinaz Koushanfar}
\email{farinaz@ucsd.edu}
\affiliation{%
  \institution{UC San Diego}
  \streetaddress{9500 Gilman Drive}
  \city{La Jolla}
  \state{CA}
  \country{USA}
  \postcode{92093}
}

\begin{abstract} 
This paper presents \sys{}, an automated and transferable watermarking framework that leverages graph neural networks to reduce the watermark search overheads during the placement stage.
\sys's novel automated watermark search is accomplished by (i) constructing novel graph and node features with physical, semantic, and design constraint-aware representation; (ii) designing a data-efficient sampling strategy for watermarking fidelity label collection; and (iii) leveraging a graph neural network to learn the connectivity between cells and predict the watermarking fidelity on unseen layouts. Extensive evaluations on ISPD’15 and ISPD’19 benchmarks demonstrate that our proposed automated methodology: (i) is capable of finding quality-preserving watermarks in a short time; and (ii) is transferable across various designs, i.e., \sys{} trained on one layout is generalizable to other benchmark circuits. \sys{} is also resilient against potential watermark removal and forging attacks.
\end{abstract}
\maketitle

\section{Introduction}
%
During the integrated circuit (IC) design flow, the physical design stage~\cite{kahng2011vlsi,sarrafzadeh1996introduction} transforms the design's logic netlist to a physical layout for fabrication. The placement~\cite{lin2019dreamplace,gnn_macroPlacerGoogle} and routing~\cite{lin2021routability,gandhi2019reinforcement} of the circuit components are optimized while ensuring various design constraints and desired functionalities. However, the IC layouts are susceptible to various security threats from the supply chain~\cite{6860363}, such as intellectual property (IP) theft, counterfeiting, and unauthorized overproduction. Watermarking~\cite{kahng2001constraint,sun2006watermarking,zhang2024icmarks} protects the physical design IP by encoding invisible, confidential, and robust signatures onto the IC layouts. These signatures help the design company to claim ownership of the physical design layouts and track the distribution of the design outcomes.

Prior physical design watermarking solutions fall into two categories~\cite{zhang2024icmarks}: (i) invasive watermarking and (ii) constraint-based watermarking. Invasive watermarking~\cite{nie2005watermarking,sun2006watermarking,bai2007watermarking,saha2007novel} adds redundant wires or cells into the layouts as watermarks. Nevertheless, the encoded signatures can be easily forged if the adversary has prior knowledge of the watermarking scheme, making it less secure for real-world applications.

Constraint-based watermarking~\cite{kahng2001constraint,zhang2024icmarks,4415879}, on the other hand, adds additional positional constraints to watermark selected cells during the placement stage. However, selecting the watermark cells without considering the modern IC design constraints, such as macros and fence regions, would significantly deteriorate quality metrics. More recent work introduced \basegp~\cite{zhang2024icmarks} to encode a watermark region of cells onto modern IC design without impacting the overall quality metrics. It aims to constrain watermark cells to be in the watermark region and the remaining cells to be out. However, searching for the ideal region constraints that adhere to design constraints and maintain watermarking fidelity consumes significant time.

This paper presents \sys{} that leverages graph neural networks (GNNs) to automate the search for the best region constraints to watermark, resulting in an efficient and transferable watermarking framework for physical design. 
By representing the layout as a graph with node features capturing physical, semantic, and design constraint information, we leverage GNNs to learn the overall quality degradation incurred by watermarking a region centered on a specific node.

\sys{} encompasses three key stages: (i) \textit{watermark search}, (ii) \textit{watermark insertion}, and (iii) \textit{watermark extraction}. During \textit{watermark search}, \sys{} employs GNN to identify the ideal region on the layout to encode the watermark. The \textit{watermark insertion} stage encodes the watermark to the cells within the watermark region during placement, similar to \cite{zhang2024icmarks}. The placement is then routed to obtain the watermarked layout. Finally, \textit{watermark extraction} employs reverse-engineering approaches~\cite{9300272,alrahis2021gnn} to acquire the cell connections and the positions from the layout. The design company can prove its ownership by decoding watermark cell positions and determining the percentage of cells within the watermark region.

In brief, our contributions are summarized as follows:

 \vspace{-5pt}
\begin{itemize}
    \item We introduce \sys, a transferable watermarking framework for physical design that leverages graph neural networks to automate the \textit{watermark search}.
    \item \sys{} preserves the watermarking fidelity by representing the design layout as a graph and employing GNN to predict the quality degradation due to watermarking at different locations in the layout.
    \item  Experiments on the ISPD'15~\cite{bustany2015ispd} and ISPD'19~\cite{liu2019ispd} benchmarks demonstrate \sys's transferability toward unseen layouts while maintaining the watermarking fidelity; It reduces the search time by 50\% for large designs ($\geq 500k$ cells) compared to ICMakrs\cite{zhang2024icmarks}. 

    \item Evaluations of \sys{} under various watermark removal and forging attacks showcase \sys's resiliency; the adversary cannot remove the signatures without significant quality degradations. 
    


\end{itemize}




\section{Background and Related Work}~\label{sec:related}
\vspace{-20pt}
\subsection{Graph Neural Networks for Physical Design}
A graph $G=(V,E)$ consists of a set of nodes $V = \{v_1, v_2,..., v_{|V|}\}$ and a set of edges $E = \{e_1, e_2,..., e_{|E|}\}$. Each node $v_i \in V$ has a $k$-dimensional feature vector $\mathbf{h}_i \in R^k$. The goal of a GNN~\cite{zhou2020graph} is to learn a function $g$ that maps the feature embedding $\mathbf{h}_i$ of node $v_i$ to a new embedding vector $\hat{\mathbf{h_i}} \in R^h$, capturing both local and global information. In a multi-layer GNN, this mapping is performed iteratively, allowing nodes to update their feature embeddings using information from their neighbors via message passing. For each node $v_i$, message passing involves receiving feature embeddings from its neighbors $j \in N_i$ and aggregating the messages using customizable functions to obtain an updated representation $\hat{\mathbf{h_i}}$. The computations are formulated in Equation~\ref{eq:GNN}, where $f$, $g$, and $\oplus$ are customizable functions, such as convolution. Here, $\mathbf{h}_v^{(l)}$ and $\mathbf{h}_u^{(l)}$ are the node features at layer $l$, $\mathcal{N}(v)$ is the set of neighbors for node $v$, and $\mathbf{h}_v^{(l+1)}$ is the updated feature of node $v$ at layer $(l+1)$.

\begin{equation}
\label{eq:GNN}
\mathbf{h}_v^{(l+1)}=g(\mathbf{h}_v^{(l)} \underset{u \in \mathcal{N}(v)}{\oplus} f(\mathbf{h}_u^{(l)}, \mathbf{h}_v^{(l)}))
\end{equation}

In physical design, graph neural networks (GNNs) serve as a promising approach during various stages of physical design, such as logic synthesis, floorplanning, partitioning, placement, and routing to improve the overall power-performance-area (PPA) metrics and speed up the chip design process~\cite{gnn_hls, gnn_powerPred, gnn_pd, gnn_3dic_partitioning, gnn_fasttuner, gnn_eda_survey, ml_eda}.
During the placement stage of physical design, GNNs have been employed to optimize for macro locations, overall timing, and power~\cite{gnn_placeOpt_clustering, gnn_macroPlacerGoogle, gnn_timingOpt, gnn_placeOpt}.
GNNs have also been employed for hardware reliability, security, and reverse engineering of gate-level netlists at the physical design stage~\cite{gnn_re, gnn_designReliabilitySecurity}.
\subsection{Physical Design Watermarking}
Physical design watermarking encodes imperceptible and unique identifiers onto the physical design layouts to protect the design company's intellectual property (IP). Invasive watermarking schemes add redundant cells or nets onto the layout as watermarks~\cite{nie2005watermarking,sun2006watermarking,bai2007watermarking,saha2007novel}, but can be easily counterfeited if the attacker has prior knowledge of the watermarking scheme. Constraint-based watermarking schemes~\cite{kahng2001constraint,kahng1998robust,ni2004watermarking,4415879} modify the cell row index and cell spacing as watermarks during the placement stage. Nevertheless, the constraint-based watermarking schemes can significantly degrade layout quality if the watermarks are selected without considering modern design constraints such as macros and fence regions.

The recent constraint-based watermarking work, ICMarks~\cite{zhang2024icmarks}, employs a scoring function to evaluate each subregion to ensure (i) sufficient cells to accommodate the signature bits; (ii) small total cell area within the watermark region for cell maneuverability; and (iii) minimal cell area overlap on the region boundary to reduce the impact of cell displacement on layout performance. ICMarks avoids regions that overlap with macros and fence regions and selects the region with the lowest score to encode the watermark. Then, the watermark insertion encodes the region of cells during placement, which co-optimizes the placement objective to ensure that only the watermark cells are in the watermark region.

\section{Motivation and Problem Formulation}~\label{sec:motivation}
\vspace{-10pt}

\textbf{Scenario:}  The IC design company establishes ownership proof for the final physical design layout by encoding watermarks during the placement stage using \sys{}.  The watermarked layout is sent along the supply chain for manufacturing and testing. To prove ownership, the design company reverse-engineers the cell locations from the layout~\cite{9300272,alrahis2021gnn} and compares the encoded watermarks with the decoded ones.

\textbf{Watermarking Criteria:} An ideal watermarking framework shall meet the following criteria:

$\bullet$ \textit{Fidelity}: The watermarked layout does not impact the functionality and the performance of the IC.

$\bullet$ \textit{Effectiveness}: The encoded watermarks can be successfully extracted along the supply chain for ownership proof. 
   
$\bullet$  \textit{Robustness}: The watermarks can withstand various removal and forging attacks and remain detectable after transformations. 

\textbf{Threats:} We consider the adversary to be a member of the supply chain, e.g., a fabrication or testing company, with access to the layout and general knowledge of watermarking algorithms. However, he/she cannot access the specific watermarking signatures or hyperparameters used by the design company. Potential threats to the watermarked layout include: 

$\bullet$ \textit{Watermark Removal Attacks}: The adversary removes the watermark by perturbing the placement solutions. 

$\bullet$  \textit{Watermark Forging Attacks}: The adversary forges the watermark by counterfeiting a different set of watermarks.

\begin{figure*}
    \centering
    \includegraphics[width= 0.9\linewidth]{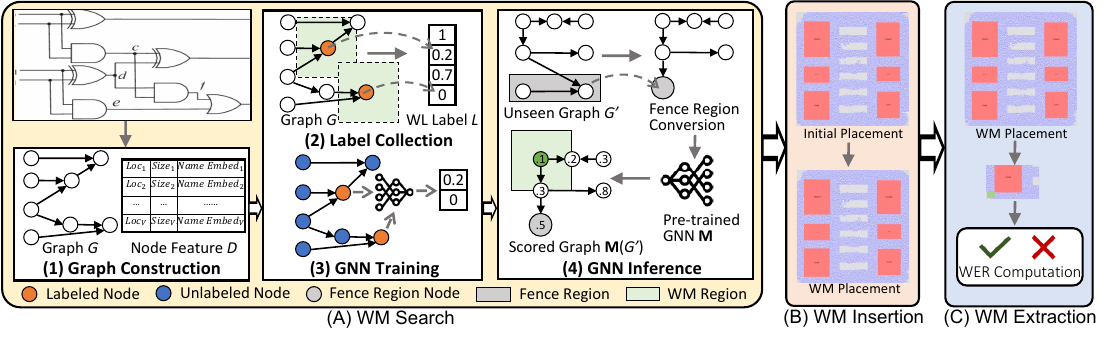}
    \vspace{-15pt}
    \caption{\sys{} flow: The \textit{Watermark Search} leverages GNN to identify the region and cells to watermark. Then, the \textit{Watermark Insertion} encodes the selected watermark region of cells on the layout during the placement stage. The \textit{Watermark Extraction} decodes the watermark and compares it with the encoded ones for ownership proof.}
    \vspace{-15pt}
    \label{fig:global}
\end{figure*}
%
\section{Methodology}~\label{sec:method}
This section introduces \sys{} and the stages in its pipeline: \textit{Watermark Search}, \textit{Watermark Insertion}, and \textit{Watermark Extraction}, as depicted in Fig.~\ref{fig:global}. \textit{Watermark Search} identifies potential regions and cells to encode the signature, and \textit{Watermark Insertion} encodes watermarks onto the IC layout without degrading the quality metrics. Finally, \textit{Watermark Extraction} decodes the watermark signature to verify ownership.
\subsection{\textit{Watermark Search}}
The objective of the \textit{Watermark Search} is to identify a region of cells that incurs minimal quality degradation after the signature is inserted. It starts by constructing a graph representing the netlist while preserving location, semantic, and design constraint features. Then, \sys{} collects the quality degradation labels on the training design and trains a graph neural network to learn the watermarking fidelity on the training nodes. The pre-trained GNN inferences on an unseen layout and the minimum-scored region node is selected as the target cell to watermark. 
\subsubsection{\textbf{Graph Construction}} The design instances, e.g., standard cells and macros, are connected to others by wired nets through the cell pins. As in Fig.~\ref{fig:global}(A-1), the design is converted into a directed homogeneous graph $G = (V, E)$, where the source nodes are the cells sending signals and the destination nodes represent the cells loading signals. The edges $E$ are the nets connecting the cells. 

For each node in the graph, we construct an 8-dimension feature $D$ as listed in Table~\ref{tab:node_embed}. The first four dimensions describe the physical information, including cell locations from initial placement and the cell size. To capture the semantic information, we extract the cell name embeddings~\cite{lu2022placement} from S-BERT~\cite{reimers2019sentence} and use Principal Component Analysis (PCA)~\cite{abdi2010principal} to reduce them to four dimensions to form the last four features of $D$.

\begin{table}[ht]
    \centering
    \vspace{-10pt}
    \begin{tabular}{|c|c|c|}
    \hline
    Type & Dim & Description \\
    \hline
    Cell Location & $2$ & from initial placement $(x, y)$ \\
    \hline
    Cell Size & $2$ &  cell size (width, height)\\
    \hline
    Cell Name & $4$ & S-BERT semantics~\cite{reimers2019sentence}\\
    \hline
     \end{tabular}
    \caption{Node Feature Construction}
    \vspace{-25pt}
    \label{tab:node_embed}
\end{table}
Modern IC designs often have additional design constraints, such as macros and fence regions. We incorporate these design constraints into the graph modeling as follows:

$\bullet$ \textit{Macros}: The macros are treated the same as other standard cells in the graph. They are distinguished from the standard cells by different cell name embeddings. 

 $\bullet$ \textit{Fence Regions}: As depicted in Fig.~\ref{fig:global}(A-4), the cells in the fence region are characterized as a single node in the graph, with cell location/size set to the fence region location/size. The name of the fence region cell is assigned after the macros.
 
\subsubsection{\textbf{Label Collection}} We collect the watermarking quality degradation labels on the \texttt{ispd19test6}~\cite{liu2019ispd} design. We chose this design as the training graph for two reasons: (i) the design is of medium size with $\sim$ 180k nodes. It ensures a complex graph structure for the GNN to learn, and the label collection takes one minute per node; (ii) the design has macros, enabling the GNN to learn the graph structure with design constraints.  The watermark region has a fixed size of $N \times row\_height$, where $N=10$.

Watermarking on every node~\footnote{For ease of explanation, we refer to watermarking the node as encoding the watermark region centered on the node throughout the paper.}  and acquiring its corresponding watermarking quality degradation would take several months for the \texttt{ispd19test6} design. Therefore, we grid the layout with the size of the watermark region and randomly sample one node from each grid as the training node. The subsampling results in approximately $3.8k$ training nodes and takes only three days to obtain all labels.  For the $i$-th node, we denote its improvement over the initial placement wirelength $WL_{INIT}$ as $\hat{L_i} = \frac{WL_i}{WL_{INIT}}$.

After acquiring the labels of each training node,  $\hat{L_i}$ is transformed into the range of $[0,1]$ for better GNN learning. As specified in Equation~\ref{eq:label}, if $\hat{L_i} < 1$, it indicates watermarking on the node will not introduce any performance degradation. Such nodes are scored as $0$, indicating an ideal position for watermark encoding. If $\hat{L_i} > \beta$, it suggests significant performance degradation from watermarking. As such, $L_i$ is set to $1$, denoting an unsatisfactory position for watermarking. For the $\hat{L_i}$ between $1$ and $\beta$, we normalize the scores between $[0,1]$. For GNN traning, we set $\beta = 1.01$. We provide additional analysis of different $\beta$ choices on \sys{} performance in Appendix~\ref{append:ablation}.
\begin{equation}
\label{eq:label}
L_i = 
\begin{cases}
    0 & \text{if } \hat{L_i} < 1 \\
    \frac{\hat{L_i}-1}{\beta} & \text{if } 1 \leq \hat{L_i} \leq \beta \\
    1 & \text{if } \hat{L_i} > \beta
\end{cases}
\end{equation}
\subsubsection{\textbf{Graph Neural Network Training}} The GNN's objective is to predict the quality metrics degradation from watermarking on the node. The model $\textbf{M}$ consists of seven-layer graph convolutions with ReLU activation~\cite{li2017convergence}. As in Equation~\ref{eq:gnn}, for the $l$-th layer, $\mathcal{N}(i)$ is the set of neighboring nodes of node $i$, $\mathbf{W}^{(l)}$ is the learnable weight matrix at layer $l$, $\mathbf{h}_j^{(l)}$ represents the feature vector of the neighboring node $j$ at layer $l$. The $\mathbf{h}_i^{(l+1)}$ is node $i$'s learned feature at layer $(l+1)$.
\begin{equation}
\label{eq:gnn}
\mathbf{h}_i^{(l+1)} = \text{ReLU}\left(\sum{j \in \mathcal{N}(i)} \frac{1}{c_{ij}} \mathbf{W}^{(l)} \mathbf{h}_j^{(l)}\right)
\end{equation}
$\textbf{M}$ is trained to minimize the MSE loss between the predicted label $\textbf{M}(G)$ and the ground truth label $L$ as in Equation~\ref{eq:mse}.  
\begin{equation}
\label{eq:mse}
\mathcal{L}_{\text{MSE}} = \frac{1}{|L|} \sum_{i=1}^{|L|} (\textbf{M}(G_i) - L_i)^2
\end{equation}
\subsubsection{\textbf{Inference on Unseen Designs}} The unseen design is converted to a graph $G^\prime$ following the \textbf{\textit{Graph Construction}}. The trained GNN $\textbf{M}$ scores each node in $G^\prime$ and obtains score $L^\prime$. The cell index $c_{wc} = \argmin(L^\prime)$ is selected as the target cell to watermark. As in Fig.~\ref{fig:global}(A-4), the watermark region  $R_w$ is centered at $c_{wc}$. The cells $C_w \in R_w$ are designated as the watermark cells.

When encoding watermarks on larger designs, where the chip canvas is much larger to accommodate more cells than the training design \texttt{ispd19test6}, a larger watermark region $R_w$ is required to ensure the watermark region is hard to remove. To address this, we perform a post-aggregation over the graph and obtain $L_{agg}^{\prime}$ using Equation~\ref{eq:agg}. The $L_{agg}^{\prime}$ better estimates the cell neighbors over larger ranges without retraining the GNN on larger designs. The target cell index is obtained as $c_{wc} = argmin(L^\prime + \gamma* L^\prime_{agg})$. We set $\gamma=0.2$ here and provide additional analysis of different $\gamma$ choices on \sys{} performance in Appendix~\ref{append:ablation}. 
\begin{equation}
\label{eq:agg}
L^{\prime}_{agg,i} = \frac{1}{N} \sum_{k=1}^{N} \left(\frac{1}{|\mathcal{N}_k(i)|} \sum_{j \in \mathcal{N}_k(i)} L^{\prime}_j\right)
\end{equation}
\subsection{\textit{Watermark Insertion}}
After obtaining the watermark region $R_w$ and the cells $C_w$, we encode them as a placement co-optimization objective to ensure only $C_w$ are placed in the region $R_w$. For a design with $K$ fence regions, \sys{} encodes an additional fence region for $R_w$ with $C_w$ as the watermark following Equation~\ref{eq:multiregion}~\cite{gu2020dreamplace}. Here, $W$ is the wirelength term, and $D$ is the cell density term with density multiplier $\lambda$. $v$ denotes the $(x,y)$ location of cell and $e\in E$ is the design net.
\begin{equation}
\label{eq:multiregion}
\begin{array}{ll}
\min _v & \sum_{e \in E} W(e ; v)+\lambda D(v), \\
\text { s.t. } & v_k=\left(x_k, y_k\right) \in R_k, \quad k=1, \cdots, R_K, R_w,
\end{array}
\end{equation}

The watermarked placement is then routed, and the design company obtains the watermarked layout.
\subsection{\textit{Watermark Extraction}}
To prove ownership, the design company reverse-engineers the suspect layout~\cite{9300272,alrahis2021gnn} and obtains netlist connectivities and cell locations. Such methods can recover large layouts containing over $7M$ cells with over $98\%$ accuracy and efficiency. Then, \sys{} queries the locations of $C_w$ and computes the cells $C_w^\prime$ within the watermark region $R_w$. The watermark extraction rate $(\%WER)$ is computed as in Equation~\ref{eq:wer}.
\begin{equation}
\label{eq:wer}
\begin{array}{ll}
\%WER = 100\times\frac{|C_w^\prime|}{|C_w|}
\end{array}
\end{equation}

\section{Experiments}~\label{sec:exp}
\vspace{-20pt}
\subsection{Experiment Setups}
\subsubsection{{Hardware Infrastructure and Implementation Details}} \sys{} is agnostic to the physical design algorithms. As a proof-of-concept, we employ the state-of-the-art physical design framework with DREAMPlace~\cite{gu2020dreamplace} as the placement algorithm and CUGR~\cite{9218646} as the routing algorithm.  
The graph neural network model is a seven-layer graph convolution network with ReLU activation. It is trained with the SGD optimizer, with the learning rate set to $0.01$, weight decay set to  $0.001$, and the momentum value set to $0.9$. It trains $30$ epochs with a batch size set to $1280$. The GNN is trained to sample 15, 20, 35, 50, 100, 200, 500 nodes at each hop. 
\subsubsection{{Benchmarks}} We evaluate \sys's watermarking performance on the ISPD'2015~\cite{bustany2015ispd}, and ISPD'2019~\cite{liu2019ispd} benchmarks. Designs with fence region constraints are highlighted in {\color[HTML]{4472C4} blue} in Tables~\ref{tab:ispd2015} and \ref{tab:ispd2019}.
\subsubsection{{Baseline}} We compare \sys{} with state-of-the-art frameworks employing both invasive and constraint-based watermarking: 

$\bullet$ \textbf{Row Parity}~\cite{kahng1998robust,kahng2001constraint} inserts unique bit sequences as watermarks by shifting cells to different rows in the placement stage. Cells with a 1-bit are moved to an odd row, while cells with a 0-bit are moved to an even row.

$\bullet$ \textbf{Cell Scattering}~\cite{4415879} scatters watermark cells on the chip canvas as watermarks using pseudorandom coordinate transformation (PRCT) algorithms. Cells with 1-bit are moved along the y-axis, and cells with 0-bit are moved along the x-axis if they do not overlap with their neighbors.

 $\bullet$ \textbf{ICMarks}~\cite{zhang2024icmarks} employs a scoring function to evaluate each layout subregion and select the best region of cells that adheres to design constraints and has minimal impact on the layout as the watermark region. The watermark region enforces only the watermark cells to be in the watermark region

  $\bullet$ \textbf{Buffer Insertion}~\cite{sun2006watermarking} adds additional buffers as watermarks during the placement stage. Two buffers are inserted to represent 0-bit, and one buffer is inserted to represent 1-bit. 
 
We skip the baselines that: (i) have different watermarking targets, e.g., smaller full-custom IC designs~\cite{ni2004watermarking,bai2007watermarking} and FPGAs~\cite{saha2007fast,liang2011chaotic}; (ii) have similar watermarking approaches as our baselines, and we use the baselines as a proof-of-concept. For example, flip-flops are encoded as watermarks in \cite{saha2007novel} instead of buffers in \cite{sun2006watermarking}.  
\subsubsection{{Evaluation Metrics}} We evaluate watermarking performance from the following metrics:

$\bullet$ \textbf{Placement WireLength Rate (PWLR)}: The rate of estimated half-perimeter wirelength (HPWL) of watermarked layout compared to the original one.

$\bullet$ \textbf{Routing WireLength Rate (RWLR)}: The rate of routed wirelength of watermarked layout compared to the original one.

$\bullet$ \textbf{Watermark Extraction Rate (WER)}: The percentage of watermark cells successfully extracted. 

 \begin{table*}[!ht]
\centering
\resizebox{0.95\textwidth}{!}{%
\begin{tabular}{|c|c|c|cc|cc|cc|cc|cc|}
\hline
& & & \multicolumn{2}{c|}{Row Parity~\cite{kahng1998robust,kahng2001constraint}} & \multicolumn{2}{c|}{\baseconstcap~\cite{4415879}}& \multicolumn{2}{c|}{\baseinvasivecap~\cite{sun2006watermarking}} & \multicolumn{2}{c|}{\basegp~\cite{zhang2024icmarks}} & \multicolumn{2}{c|}{\sys}\\ \cline{4-13} 
\multirow{-2}{*}{Design} & \multirow{-2}{*}{Cells} &  \multirow{-2}{*}{Nets} &  \multicolumn{1}{c|}{PWLR $\downarrow$} & RWLR $\downarrow$ & \multicolumn{1}{c|}{PWLR $\downarrow$} & RWLR $\downarrow$ & \multicolumn{1}{c|}{PWLR $\downarrow$} & RWLR $\downarrow$ & \multicolumn{1}{c|}{PWLR $\downarrow$} & RWLR $\downarrow$ & \multicolumn{1}{c|}{PWLR $\downarrow$} & RWLR $\downarrow$ \\ \hline
\hline 
{\color[HTML]{4472C4} perf\_a $\star$}& 108K & 115K & \multicolumn{1}{c|}{1.0045} & 1.0155 &\multicolumn{1}{c|}{0.9978} & 0.9967 & \multicolumn{1}{c|}{\textcolor{gray}{1.5289}} &  \textcolor{gray}{2.3270} & \multicolumn{1}{c|}{0.9972} & \textbf{0.9873} & \multicolumn{1}{c|}{\textbf{0.9956}} & 0.9929\\ \hline
{\color[HTML]{4472C4} perf\_b} & 113K & 113K & \multicolumn{1}{c|}{1.0058} & 1.0267&  \multicolumn{1}{c|}{1.0020} & 1.0001 & \multicolumn{1}{c|}{1.0176} & 1.0745 & \multicolumn{1}{c|}{\textbf{0.9890}} &0.9901 & \multicolumn{1}{c|}{0.9898} &  \textbf{0.9824}\\ \hline
{\color[HTML]{4472C4} dist\_a} & 127K & 134K  & \multicolumn{1}{c|}{1.0015} & 0.9999 & \multicolumn{1}{c|}{\textbf{0.9984}} & \textbf{0.9965} & \multicolumn{1}{c|}{1.0995} & 1.1072  & \multicolumn{1}{c|}{1.0004} & 1.0052 & \multicolumn{1}{c|}{1.0001} & 1.0040\\ \hline
{\color[HTML]{4472C4} mult\_b $\star$} & 146K & 152K & \multicolumn{1}{c|}{1.0047} & 1.0199 & \multicolumn{1}{c|}{\textbf{0.9991}}  & 0.9923& \multicolumn{1}{c|}{\textcolor{gray}{1.8966}} & \textcolor{gray}{2.9374}  & \multicolumn{1}{c|}{0.9994} & 1.0028 &  \multicolumn{1}{c|}{1.0021} & \textbf{0.9922}\\ \hline
{\color[HTML]{4472C4} mult\_c $\star$} & 146K & 152K & \multicolumn{1}{c|}{1.0031} & 1.0252 & \multicolumn{1}{c|}{0.9980}& 
1.0023& \multicolumn{1}{c|}{\textcolor{gray}{1.6003}} & \textcolor{gray}{2.7459}  &  \multicolumn{1}{c|}{\textbf{0.9963}} & \textbf{0.9979} & \multicolumn{1}{c|}{1.0010} & 1.0022\\ \hline
{\color[HTML]{4472C4} pci\_a $\star$} & 30K & 34K & \multicolumn{1}{c|}{1.0080} & 1.0340 & \multicolumn{1}{c|}{\textbf{0.9931}} &\textbf{ 0.9846} & \multicolumn{1}{c|}{\textcolor{gray}{1.6269}} & \textcolor{gray}{1.6849}& \multicolumn{1}{c|}{0.9997} & 0.9950 & \multicolumn{1}{c|}{0.9996} & 0.9974\\ \hline
{\color[HTML]{4472C4} pci\_b $\star$}  & 29K & 33K  & \multicolumn{1}{c|}{1.0069} & 1.1173 &\multicolumn{1}{c|}{\textbf{0.9912}}  & 0.9977 & \multicolumn{1}{c|}{\textcolor{gray}{1.2239}} & \textcolor{gray}{1.8207} & \multicolumn{1}{c|}{0.9951} & 1.0026 & \multicolumn{1}{c|}{0.9973} & \textbf{0.9885}\\ \hline
{\color[HTML]{4472C4} superblue11 $\star$} & 926K & 936K& \multicolumn{1}{c|}{1.0052} & 1.0344 &  \multicolumn{1}{c|}{1.0278} & 1.0358 & \multicolumn{1}{c|}{\textcolor{gray}{1.6930}} & \textcolor{gray}{3.7086} & \multicolumn{1}{c|}{\textbf{0.9986}} & \textbf{0.9992} & \multicolumn{1}{c|}{1.0073} & 1.0136 \\ \hline
{\color[HTML]{4472C4} superblue16} & 680K & 697K  & \multicolumn{1}{c|}{1.0030} & 1.0210 & \multicolumn{1}{c|}{\textbf{0.9988}} & \textbf{0.9989} & \multicolumn{1}{c|}{1.1023} & 1.1377 & \multicolumn{1}{c|}{1.0017} & 1.0204 & \multicolumn{1}{c|}{1.0025} & 1.0129\\ \hline
perf\_1 & 113K & 113K  & \multicolumn{1}{c|}{1.0035} & 1.0199 & \multicolumn{1}{c|}{0.9985} & 0.9983 & \multicolumn{1}{c|}{1.0150} &  1.0642 &  \multicolumn{1}{c|}{0.9964} & 0.9973 & \multicolumn{1}{c|}{\textbf{0.9909}} & \textbf{0.9956} \\ \hline
fft\_1 &  35K & 33K & \multicolumn{1}{c|}{1.0017} & 1.0260 & \multicolumn{1}{c|}{1.0167} & 1.0156 & \multicolumn{1}{c|}{1.0680} &1.1457 &  \multicolumn{1}{c|}{\textbf{0.9671}} & 0.9673 & \multicolumn{1}{c|}{0.9711} & \textbf{0.9660}\\ \hline
fft\_2 $\star$ & 35K & 33K & \multicolumn{1}{c|}{1.0050} & 1.0260 & \multicolumn{1}{c|}{1.0148}& 1.0114 & \multicolumn{1}{c|}{\textcolor{gray}{1.4603}} & \textcolor{gray}{1.4476}  &   \multicolumn{1}{c|}{\textbf{0.9767}} & \textbf{0.9770} & 0.9854 & 0.9823\\ \hline
fft\_a $\star$ & 34K & 32K & \multicolumn{1}{c|}{1.0121} & 1.0200 & \multicolumn{1}{c|}{1.0024}& 0.9933& \multicolumn{1}{c|}{\textcolor{gray}{1.8203}} & \textcolor{gray}{2.1923}  & \multicolumn{1}{c|}{0.9939} & 0.9895 & \multicolumn{1}{c|}{\textbf{0.9839}} & \textbf{0.9773} \\ \hline
fft\_b $\star$& 34K & 32K& \multicolumn{1}{c|}{1.0018} & 1.0075 & \multicolumn{1}{c|}{0.9961} &1.0030 & \multicolumn{1}{c|}{\textcolor{gray}{1.9859}} & \textcolor{gray}{2.5615} &  \multicolumn{1}{c|}{0.9909} & \textbf{0.9890} & \multicolumn{1}{c|}{\textbf{0.9869}} & 1.0039 \\ \hline
mult\_1 & 160K & 159K &\multicolumn{1}{c|}{1.0050} & 1.0238 & \multicolumn{1}{c|}{1.0077} & 1.0065 & \multicolumn{1}{c|}{1.0464} &1.0631  &  \multicolumn{1}{c|}{\textbf{0.9753}} & \textbf{0.9744} &  \multicolumn{1}{c|}{0.9773} & 0.9769\\ \hline
mult\_2  & 160K & 159K & \multicolumn{1}{c|}{1.0033} & 1.0199 &  \multicolumn{1}{c|}{0.9984} & 0.9967 & \multicolumn{1}{c|}{1.0736} & 1.1147&  \multicolumn{1}{c|}{0.9852} & 0.9900 & \multicolumn{1}{c|}{\textbf{0.9831}} & \textbf{0.9872}\\ \hline
mult\_a $\star$ & 154K & 154K& \multicolumn{1}{c|}{1.0037} & 1.0105 & \multicolumn{1}{c|}{ 0.9995}& 0.9968 & \multicolumn{1}{c|}{\textcolor{gray}{1.3862}} & \textcolor{gray}{1.6738}  &\multicolumn{1}{c|}{0.9973} & \textbf{0.9916} & \multicolumn{1}{c|}{\textbf{0.9965}} & 0.9934\\ \hline
superblue12 & 1293K & 1293K& \multicolumn{1}{c|}{1.0031} & 1.0067 &\multicolumn{1}{c|}{0.9979} & 0.9956 & \multicolumn{1}{c|}{1.0683} & 1.1044 & \multicolumn{1}{c|}{\textbf{0.9854}} & \textbf{0.9732} & \multicolumn{1}{c|}{0.9869} & 0.9783\\ \hline
superblue14 & 634K & 620K & \multicolumn{1}{c|}{1.0020} & 1.0057 & \multicolumn{1}{c|}{0.9991} & 0.9981 & \multicolumn{1}{c|}{1.0212} & 1.0286 & \multicolumn{1}{c|}{\textbf{0.9887}} & \textbf{0.9867} & \multicolumn{1}{c|}{1.0083} & 1.0037 \\ \hline
superblue19 & 522K & 512K & \multicolumn{1}{c|}{1.0025} & 1.0077 & \multicolumn{1}{c|}{1.0001} & 1.0005& \multicolumn{1}{c|}{1.0295} & 1.0672&  \multicolumn{1}{c|}{\textbf{0.9814}} & \textbf{0.9809} & \multicolumn{1}{c|}{1.0100} & 1.0563\\ \hline
\hline 
Average & - & - & \multicolumn{1}{c|}{1.0043} & 1.0231 & \multicolumn{1}{c|}{1.0018} & 1.0012 & \multicolumn{1}{c|}{1.0536} & 1.0901&  \multicolumn{1}{c|}{\textbf{0.9908}} &\textbf{0.9908} & \multicolumn{1}{c|}{0.9937} &  0.9966 \\ \hline
\end{tabular}%
}
\caption{Performance on the ISPD'2015 benchmarks~\cite{bustany2015ispd}. All the design watermarks are successfully extracted, i.e., WER = 100\%. The PWLR and RWLR are the placement and routed wirelength rates over the original designs.  The results in \textcolor{gray}{gray} fail buffer insertion WM with significant degradation on the high-utilized designs (denoted with $\star$).
}
\vspace{-20pt}
\label{tab:ispd2015}
\end{table*}

\begin{table*}[!ht]
\centering
\resizebox{0.95\textwidth}{!}{%
\begin{tabular}{|c|c|c|cc|cc|cc|cc|cc|}
\hline
& & & \multicolumn{2}{c|}{Row Parity~\cite{kahng1998robust,kahng2001constraint}} & \multicolumn{2}{c|}{\baseconstcap~\cite{4415879}}& \multicolumn{2}{c|}{\baseinvasivecap~\cite{sun2006watermarking}} & \multicolumn{2}{c|}{\basegp~\cite{zhang2024icmarks}} & \multicolumn{2}{c|}{\sys}\\ \cline{4-13} 
\multirow{-2}{*}{Design} & \multirow{-2}{*}{Cells} &  \multirow{-2}{*}{Nets} &  \multicolumn{1}{c|}{PWLR $\downarrow$} & RWLR $\downarrow$ & \multicolumn{1}{c|}{PWLR $\downarrow$} & RWLR $\downarrow$ & \multicolumn{1}{c|}{PWLR $\downarrow$} & RWLR $\downarrow$ & \multicolumn{1}{c|}{PWLR $\downarrow$} & RWLR $\downarrow$ & \multicolumn{1}{c|}{PWLR $\downarrow$} & RWLR $\downarrow$ \\ \hline
\hline 
ispd19test1 & 9K & 3K & \multicolumn{1}{c|}{1.0060} & 1.0129 & \multicolumn{1}{c|}{0.9978} & 0.9998 & \multicolumn{1}{c|}{1.0394} & 1.0619 & \multicolumn{1}{c|}{\textbf{0.9955}} & 1.0015  & \multicolumn{1}{c|}{1.0026} &\textbf{0.9989} \\ \hline
ispd19test2 & 73K & 72K &  \multicolumn{1}{c|}{1.0184} & 1.0201 & \multicolumn{1}{c|}{1.0096} & 1.0016 & \multicolumn{1}{c|}{1.0127} & 1.0408 &\multicolumn{1}{c|}{0.9988} & 0.9999  & \multicolumn{1}{c|}{\textbf{0.9921}} & \textbf{0.9927} \\ \hline
ispd19test3 & 8K & 9K & \multicolumn{1}{c|}{1.0130} & 1.0475 & \multicolumn{1}{c|}{1.0180} & 1.0193 & \multicolumn{1}{c|}{1.0582} & 1.0801 &\multicolumn{1}{c|}{1.0059} & 1.0045 & \multicolumn{1}{c|}{\textbf{0.9875}} &  \textbf{0.9865}\\ \hline
ispd19test4 & 151K & 146K & \multicolumn{1}{c|}{1.0017} & 1.0050 & \multicolumn{1}{c|}{0.9999} & 0.9998 & \multicolumn{1}{c|}{1.1452} & 1.2494 & \multicolumn{1}{c|}{ \textbf{0.9957}} &  \textbf{0.9907} &\multicolumn{1}{c|}{ 0.9970}& 1.0001 \\ \hline
{\color[HTML]{4472C4} ispd19test5} & 29K & 29K & \multicolumn{1}{c|}{1.0102} & 1.0556 & \multicolumn{1}{c|}{1.0998} & 1.0975 & \multicolumn{1}{c|}{0.9859}  & 1.0121 & \multicolumn{1}{c|}{1.0013} & 0.9998 & \multicolumn{1}{c|}{\textbf{0.9982}} & \textbf{0.9953}\\ \hline
ispd19test6 & 181K & 180K & \multicolumn{1}{c|}{1.0032} & 1.0106 & \multicolumn{1}{c|}{\textbf{0.9994}} & 0.9993 & \multicolumn{1}{c|}{1.0044} & 1.1108 &  \multicolumn{1}{c|}{1.0023} & 1.0026  & \multicolumn{1}{c|}{1.0000} & \textbf{0.9971} \\ \hline
ispd19test7 & 362K & 359K & \multicolumn{1}{c|}{1.0028} & 1.0160 & \multicolumn{1}{c|}{1.0136} & 1.0109 & \multicolumn{1}{c|}{\textbf{1.0003}} & 1.0717 & \multicolumn{1}{c|}{1.0050} & 1.0054 & \multicolumn{1}{c|}{1.0150} & 1.0139 \\ \hline
ispd19test8 & 543K & 538K & \multicolumn{1}{c|}{1.0001} & 1.0082 &  \multicolumn{1}{c|}{0.9966} & 0.9958 & \multicolumn{1}{c|}{1.0118} & 1.0806 & \multicolumn{1}{c|}{0.9961} & \textbf{0.9929} &\multicolumn{1}{c|}{\textbf{0.9945}} &0.9996  \\ \hline
ispd19test9 & 903K & 895K & \multicolumn{1}{c|}{1.0072} & 1.0108 & \multicolumn{1}{c|}{1.0108} & 1.0095 & \multicolumn{1}{c|}{1.0164} & 1.0814 & \multicolumn{1}{c|}{1.0023} & 1.0023 & \multicolumn{1}{c|}{\textbf{0.9973}} & \textbf{1.0000} \\ \hline
ispd19test10 & 903K & 895K & \multicolumn{1}{c|}{1.0025} & 1.0090 & \multicolumn{1}{c|}{1.0043} & 1.0108 & \multicolumn{1}{c|}{1.0378} & 1.0982 &\multicolumn{1}{c|}{0.9972} & \textbf{0.9966} & \multicolumn{1}{c|}{\textbf{0.9972}} & 1.0002 \\ \hline
\hline 
Average & - & - & \multicolumn{1}{c|}{1.0065} & 1.0194 & \multicolumn{1}{c|}{1.0060}& 1.0140 & \multicolumn{1}{c|}{1.0304} &1.0871 & \multicolumn{1}{c|}{1.0000} & 0.9996 & \multicolumn{1}{c|}{\textbf{0.9981}} & \textbf{0.9982} \\ \hline
\end{tabular}%
}
\caption{Performance on the ISPD'2019 benchmarks~\cite{liu2019ispd}. All the design watermarks are successfully extracted, i.e., WER = 100\%. The PWLR and RWLR are the placement and routed wirelength rates over the original designs.}
\vspace{-15pt}
\label{tab:ispd2019}
\end{table*}
\subsection{\sys{} Results}\label{sec:results}
\subsubsection{\textbf{Fidelity and Transferability}} We compare the watermarking performance of \sys{} and the baselines on the ISPD'2015~\cite{bustany2015ispd} benchmarks in Table~\ref{tab:ispd2015}, and the ISPD'2019~\cite{liu2019ispd} benchmarks in Table~\ref{tab:ispd2019}. The encoded watermarks are extracted successfully, i.e., $\%WER = 100$ for all frameworks.
We highlight that the \sys{} is only trained on the \texttt{ispd19test6} design and inference on the rest of the designs, significantly reducing the watermark search time and improving transferability.  

\textbf{Comparison with Constraint-based WM:} 
Compared to \sys{} that maintains watermarking fidelity, the baseline Row Parity~\cite{kahng1998robust,kahng2001constraint} and  \baseconstcap~\cite{4415879} degrades the PWLR by $0.18\%$ and $0.60\%$ and the RWLR by $0.12\%$ and $1.40\%$ over the non-wm designs on ISPD'2015~\cite{bustany2015ispd} and ISPD'2019~\cite{liu2019ispd} benchmarks respectively.
The Row Parity~\cite{kahng1998robust,kahng2001constraint} and \baseconstcap~\cite{4415879} approaches do not consider the design constraints, like fence regions or macros when selecting the watermark cells, resulting in performance degradation.

Compared to \sys{}, \basegp~\cite{zhang2024icmarks} introduced slightly more degradations on ISPD'2019~\cite{liu2019ispd} benchmarks. It is because \sys{} learns the mapping from layout subgraphs to the actual PWLR improvement, which serves a better quality degradation estimation than the scoring function used in \basegp~\cite{zhang2024icmarks}. 

\textbf{Comparison with Invasive WM:} Compared to \sys, the invasive watermarking \baseinvasivecap~\cite{sun2006watermarking} significantly degrades the PWLR and RWLR by $5.36\%$ and $9.01\%$ on the ISPD'2015~\cite{bustany2015ispd} designs; and by $3.04\%$ and $8.71\%$ on the ISPD'2019~\cite{liu2019ispd} benchmarks over the non-watermarked designs. As the layout is highly utilized, adding redundant buffers results in significant cell placement reordering to accommodate the watermark cells, requiring more routing resources to connect the additional components than constraint-based \sys.
\subsubsection{\textbf{Efficiency}}
\begin{figure}[ht!]
\vspace{-10pt}
    \begin{minipage}[c]{0.7\columnwidth}
       \includegraphics[width=\columnwidth]{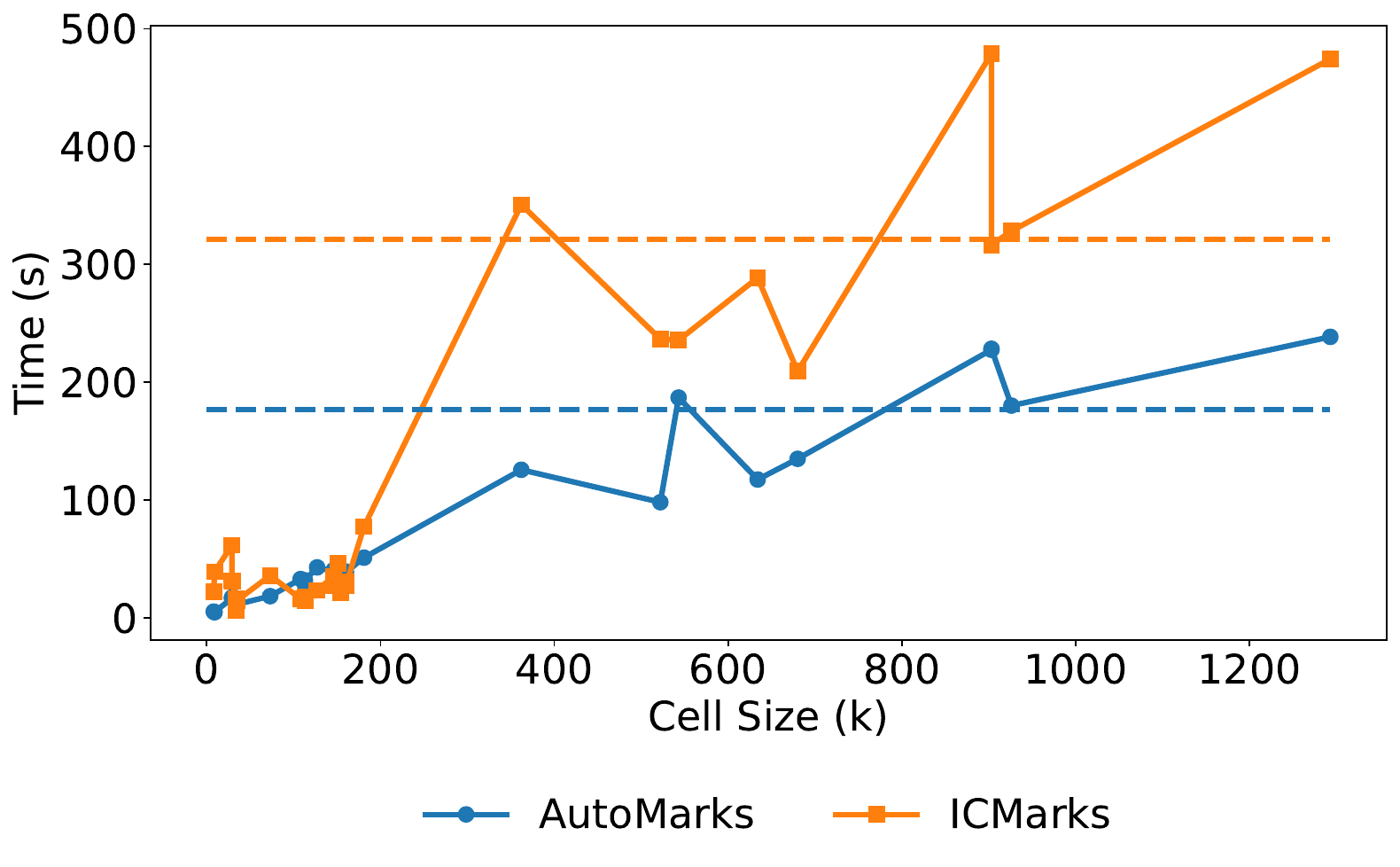}
    \end{minipage}
    \hfill
    \begin{minipage}[c]{0.29\columnwidth}
       \caption{\small{The watermark search time for different designs. The dotted line is the average search time of the large designs ($\geq 500k$ cells).}}
        \label{fig:efficiency}
    \end{minipage}
    \vspace{-20pt}
\end{figure}

Fig.~\ref{fig:efficiency} compares the watermark search time for \basegp~\cite{zhang2024icmarks} and \sys. For smaller designs, the search time of \sys{} is very close to \basegp~\cite{zhang2024icmarks}. However, for the large designs($\geq 500k$ cells), the average search time of \sys{} is $176.38s$, whereas \basegp{} requires $320.93s$. As \basegp{} uses a fixed window size to traverse and score the layout, while \sys{} employs GNN to batch-predict the region node scores, \sys{} reduces the watermark search time for large designs by $45.04\%$.
%
%
%
%

%
\subsection{\sys's Robustness}
%
%
\subsubsection{\textbf{Watermark Removal Attacks}}
\begin{figure*}[!ht]
    \begin{minipage}[c]{0.78\textwidth}
       \subfloat[Location swap attack (0.1\%) ]{\label{f:sla_10_attack} \includegraphics[width=0.5\linewidth]{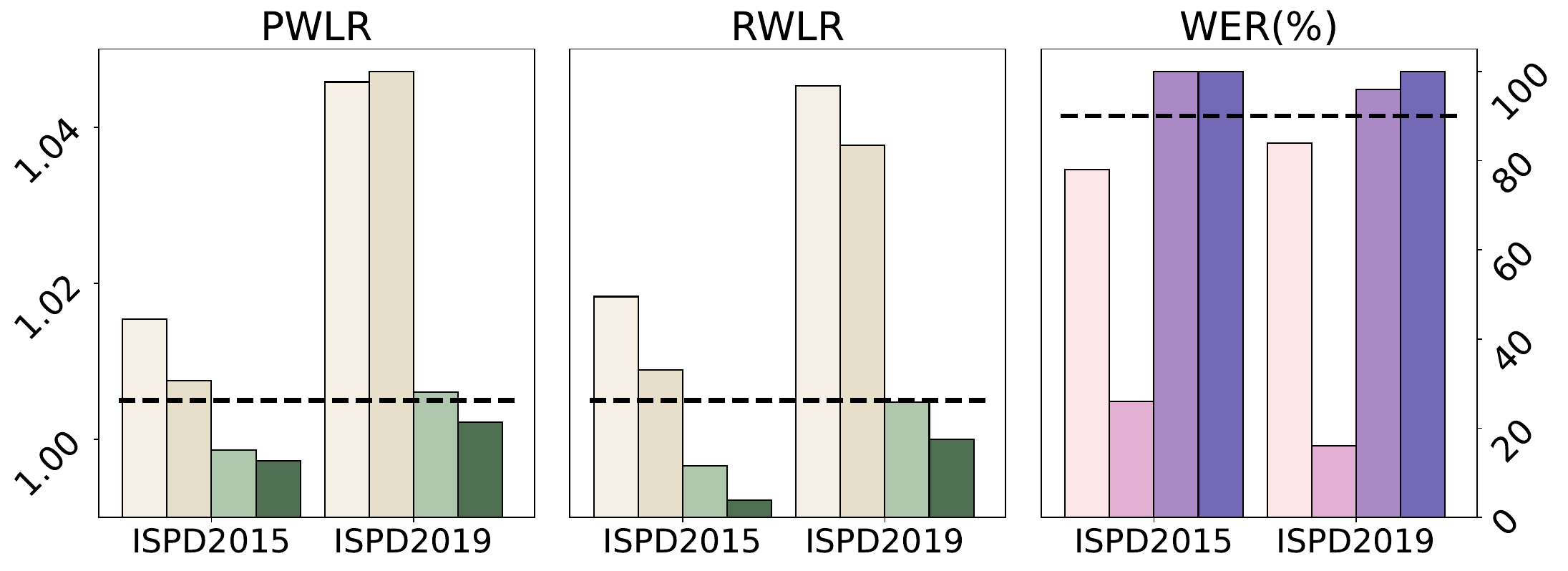}}
    \subfloat[Constraint perturbation attack (10\%)]{\label{f:cpa_0.1} \includegraphics[width=0.5\linewidth]{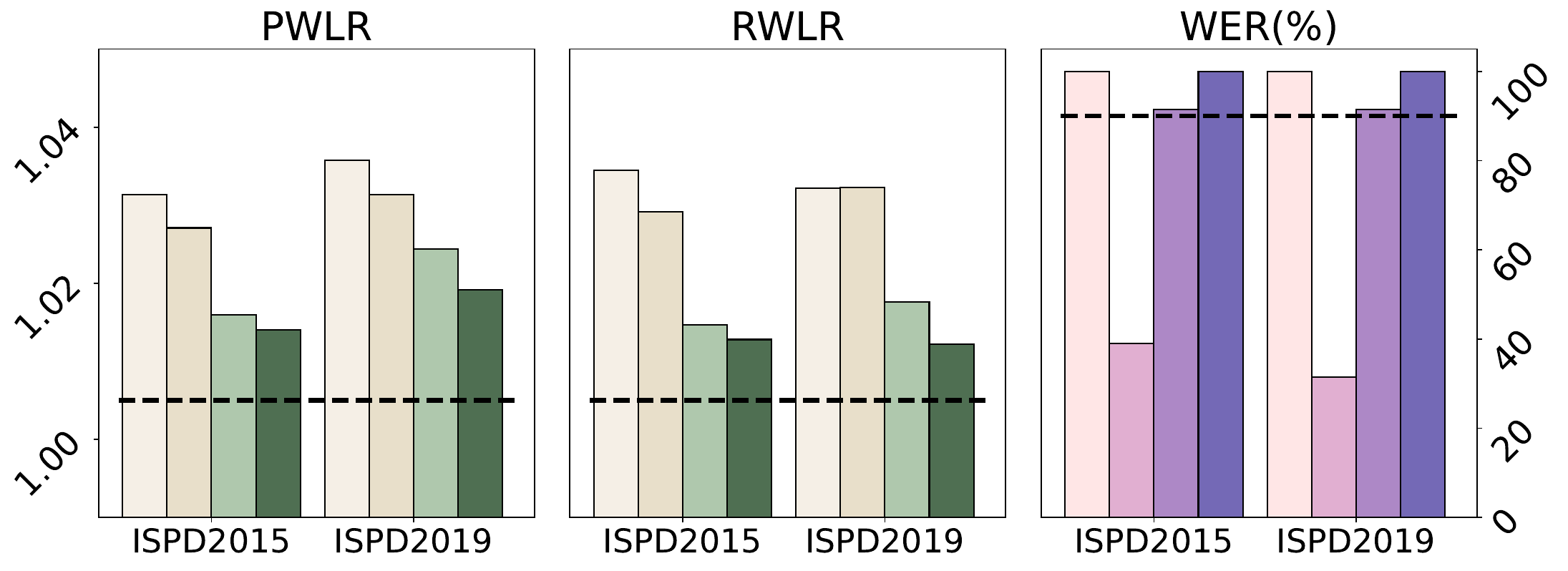}}  \\
    \subfloat[Optimization attack ]{\label{f:eco_attack} 
    \includegraphics[width=0.5\linewidth]{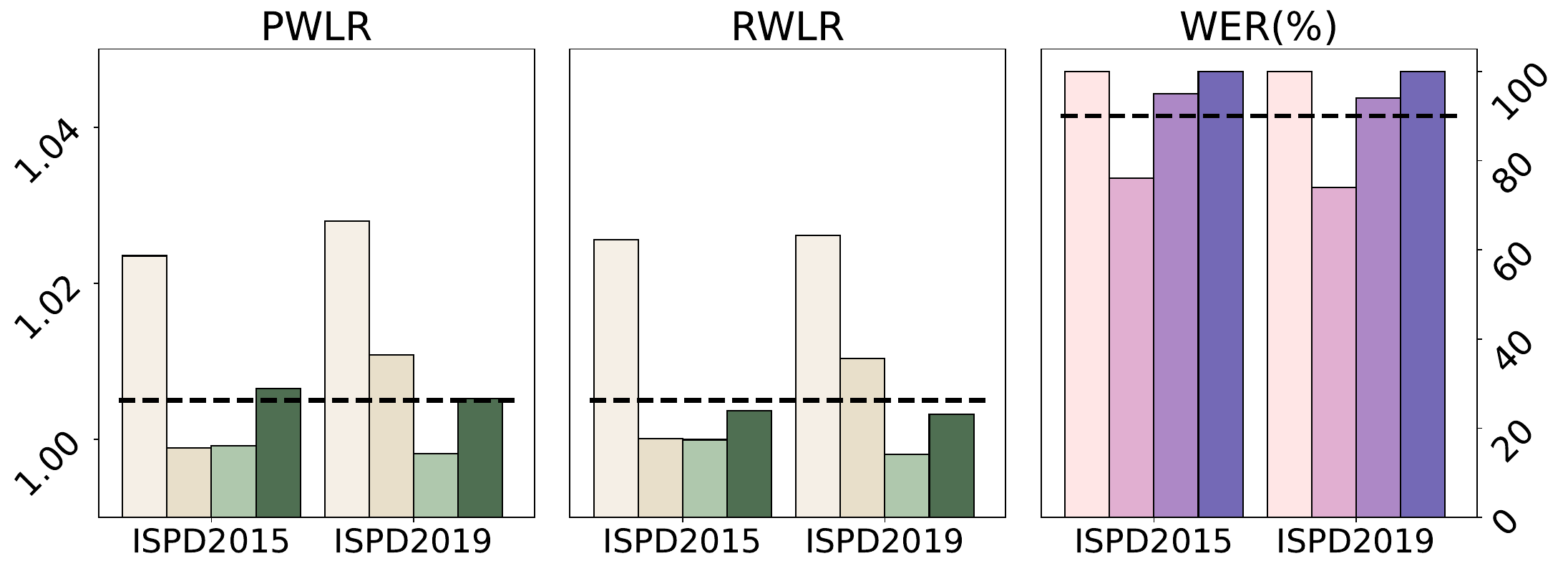}} 
   \subfloat[Adaptive region attack (top-5) ]{\label{f:ada_attack5} \includegraphics[width=0.5\linewidth]{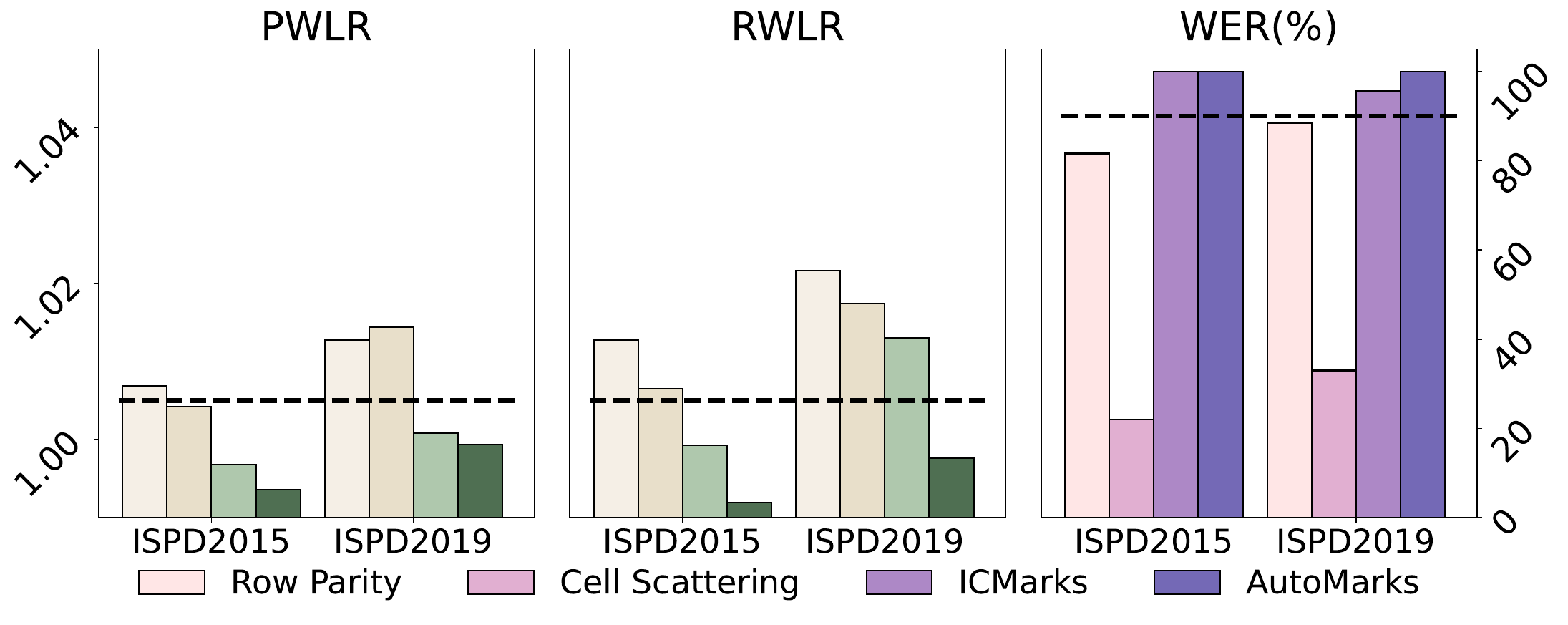}}
    \end{minipage}
    \hfill
    \begin{minipage}[c]{0.19\textwidth}
        \caption{\small{Watermarking performance under different attacks for wirelength-driven placement on the ISPD'2015~\cite{bustany2015ispd} and ISPD'2019~\cite{liu2019ispd} benchmarks. The black dotted line in the two left subfigures denotes the quality degradation threshold of 1.005, and the black dotted line in the rightmost subfigure denotes the watermark extraction threshold of 90\%}.}
        \label{fig:attack_results1}
    \end{minipage}
    \vspace{-6pt}
\end{figure*}

Fig.~\ref{fig:attack_results1} evaluates the impact of watermark removal attacks on \sys{} and the baseline approaches. Due to space limitation, more details of the attacks under different parameter settings are in Appendix~\ref{append:attack}. The \textbf{Location swap attack} targeting Row Parity~\cite{kahng1998robust,kahng2001constraint} selects 0.1\% of the total cells randomly and swaps their locations. The \textbf{Constraint perturbation attack} targeting Cell Scattering~\cite{4415879} moves $10\%$ of the cells randomly along the x-axis for one unit or the y-axis for one-row height if the cells do not overlap with the neighboring cells.
The \textbf{Optimization attack} targeting all frameworks applies another round of detailed placement on the watermarked layout to remove signatures.
The \textbf{Adaptive region attack} targeting ICMarks~\cite{zhang2024icmarks} and \sys, assumes the adversary knows the size of the watermark region and uses the evaluation function of \basegp{}~\cite{zhang2024icmarks} to identify the watermark regions. Then, the adversary perturbs cells in the top-5 regions to remove watermarks. We do not consider attacks on \baseinvasivecap{} as it incurs significantly more performance degradation for watermark insertion than the constraint-based watermarking approaches.  

\sys{} is resilient to all attacks and maintains the watermark extraction rates of 100\% even when the quality metrics PWLR and RWLR are greatly compromised; because as long as the watermark cells are within the watermark region, it is more robust toward such slight perturbation of cell locations.
In contrast, Row Parity is vulnerable to Location swap attacks; \baseconstcap{} is vulnerable to Location swap attacks and Constraint perturbation attacks. While \basegp{} employs a two-level watermarking framework to strengthen the watermarking strength, the detailed watermarking in \basegp{} is vulnerable to watermark removal attacks. As a result, the overall WER for \basegp{} slightly degrades. 
%
%
%
%
%
%
%
\subsubsection{\textbf{Watermark Forging Attacks}} To forge the signature, the adversary needs to provide the watermark region by reproducing the GNN inference results. However, the adversary does not have access to the design used for GNN training, making it difficult to counterfeit the signature. Therefore, \sys{} is resilient to the forging attacks. 
%
\subsection{Ablation Study and Analysis}\label{sec:ablation}
This subsection provides ablation studies over the different hyperparameter choices in \sys's performance using the ISPD'2019~\cite{liu2019ispd} as benchmarks. Additional visualizations are in Appendix~\ref{append:visual}, and ablation studies are in Appendix~\ref{append:ablation}.
\subsubsection{\textbf{The Effectiveness of Node Feature}} We analyze how different node features impact the GNN performance. In Table~\ref{tab:feature}, we skip the cell location, size, and name as the node feature and report their PWLR and RWLR performance respectively. Other settings follow the default ones in \sys. 

Cell location and name have a more significant impact on the performance of the \sys{} compared to cell size. Excluding the cell location and name into the node feature construction results in 2.9\% and 3.2\% PWLR and 3.4\% and 3.3\% RWLR degradation, respectively. In contrast, excluding the cell size only results in 0.2\% degradation. This is mainly because the cell name indicates the type of cells, e.g., standard cells and macros, and cell location indicates the position of the cells. Both attributes are essential in helping \sys{} learn which node is the ideal candidate for watermark insertion. 
\begin{table}[ht]
\vspace{-10pt}
    \centering
    \begin{tabular}{|c|c|c|}
    \hline
        Node Feature & PWLR & RWLR \\\hline
       All & 0.9981 & 0.9982 \\
       w/o cell location &1.0280 &1.0312\\ 
       w/o cell size & 1.0002& 0.9977\\
       w/o cell name & 1.0303& 1.0304\\ \hline
    \end{tabular} 
    \caption{The effectiveness of different node feature constructions on \sys's performance.}
    \vspace{-20pt}
    \label{tab:feature}
\end{table}

%

\vspace{-10pt}
\section{Conclusion}~\label{sec:con}



This work presents \sys, an automated and highly transferable watermarking framework for physical design. Leveraging graph neural networks for the watermark region search, \sys{} significantly reduces the search time while preserving watermarking fidelity. Extensive evaluations on the ISPD'15 and ISPD'19 benchmarks demonstrate the effectiveness and transferability of our proposed framework compared to existing physical design watermarking approaches. \sys{} is resilient against watermarking removal and forging attacks with $100\%$ watermark extraction rate for proof of ownership.

 
\begin{acks}

This work was supported by NSF TILOS AI Institute award number 2112665.

\end{acks}

\bibliographystyle{ACM-Reference-Format}
\bibliography{bib}

\onecolumn
\appendix
\subsection{Additional Attack Evaluations}~\label{append:attack}
In this subsection, we include the watermark removal attack evaluations with different parameter settings, including location swap attack, constraint perturbation attack, and adaptive region attack.
%
\begin{figure*}[ht]
    \centering
    \subfloat[Location swap attack (0.5\%)]{\label{f:sla_50_attack} \includegraphics[width=0.5\linewidth]{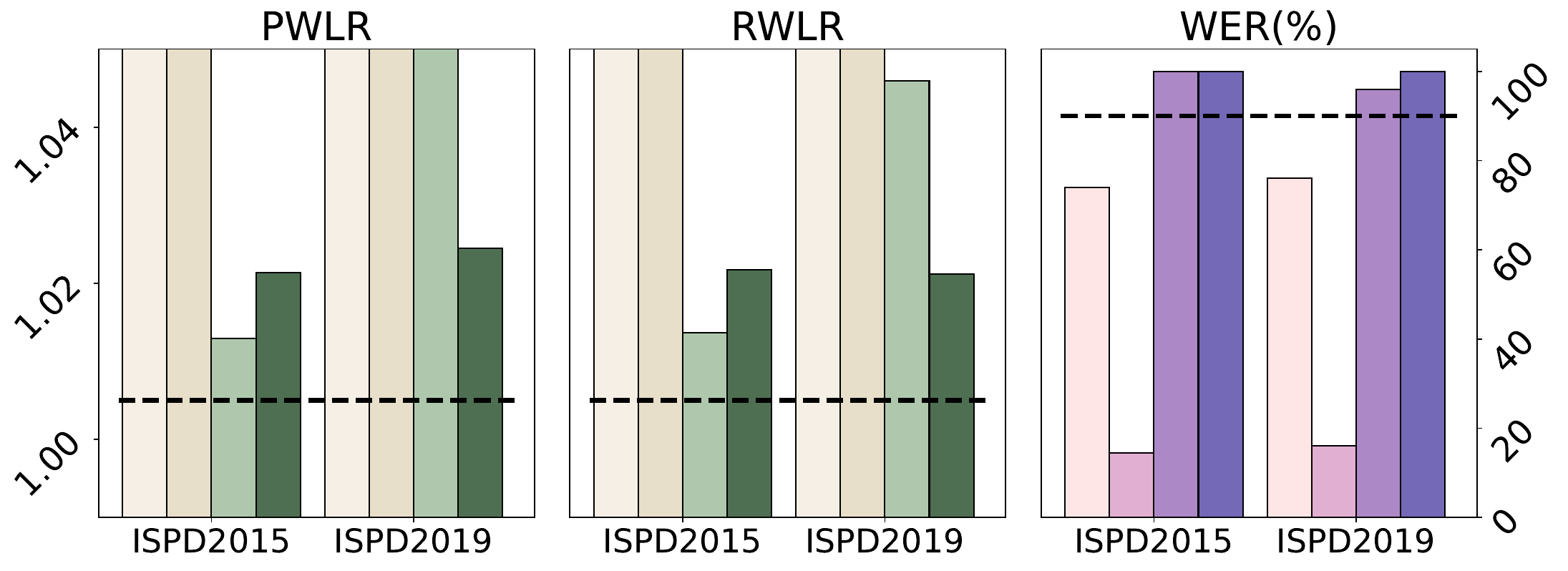}}  
     \subfloat[Constraint perturbation attack (0.1\%) ]{\label{f:cpa_0.001} \includegraphics[width=0.5\linewidth]{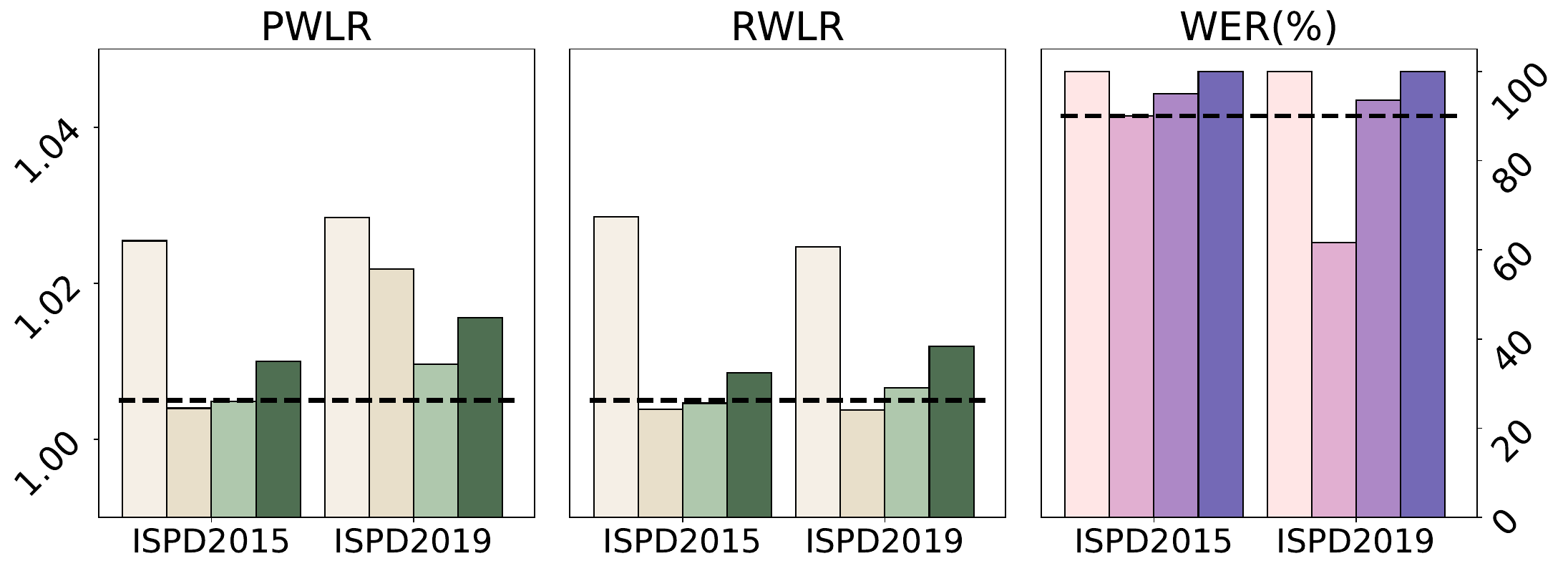}} \\ 
    \subfloat[Constraint perturbation attack (1\%) ]{\label{f:cpa_0.01} \includegraphics[width=0.5\linewidth]{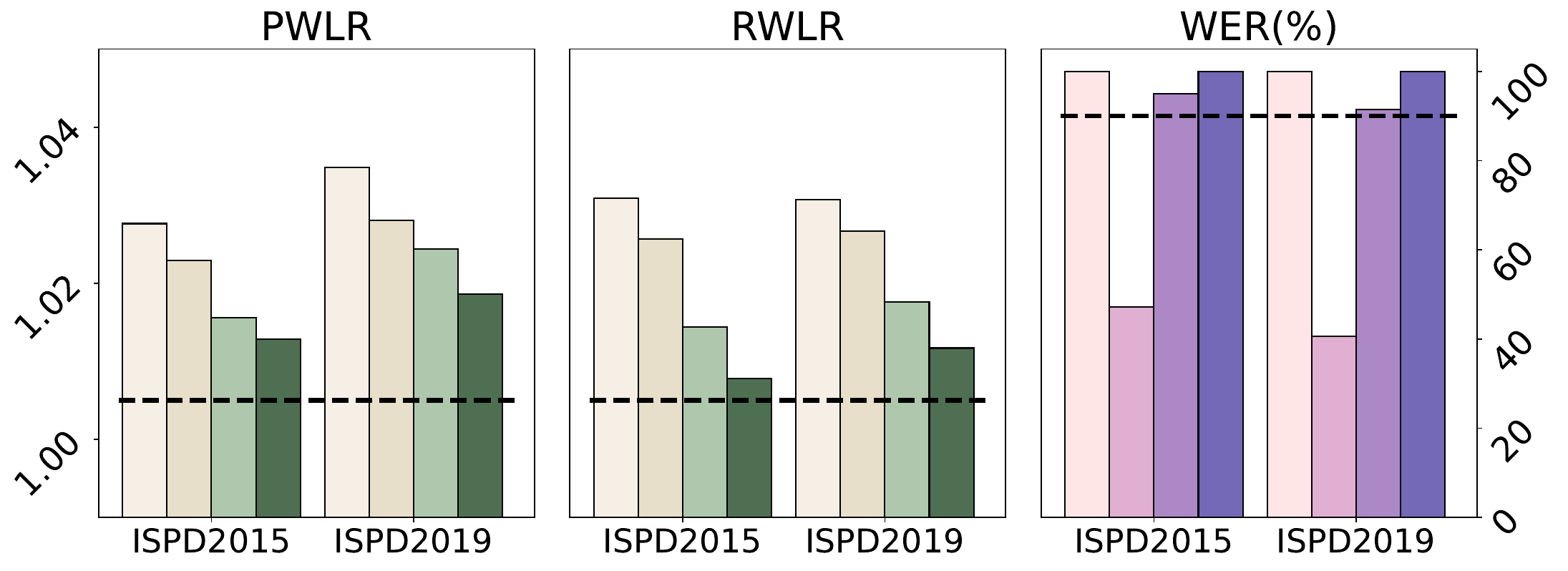}} 
    \subfloat[Adaptive region attack (top-1) ]{\label{f:ada_attack1} \includegraphics[width=0.5\linewidth]{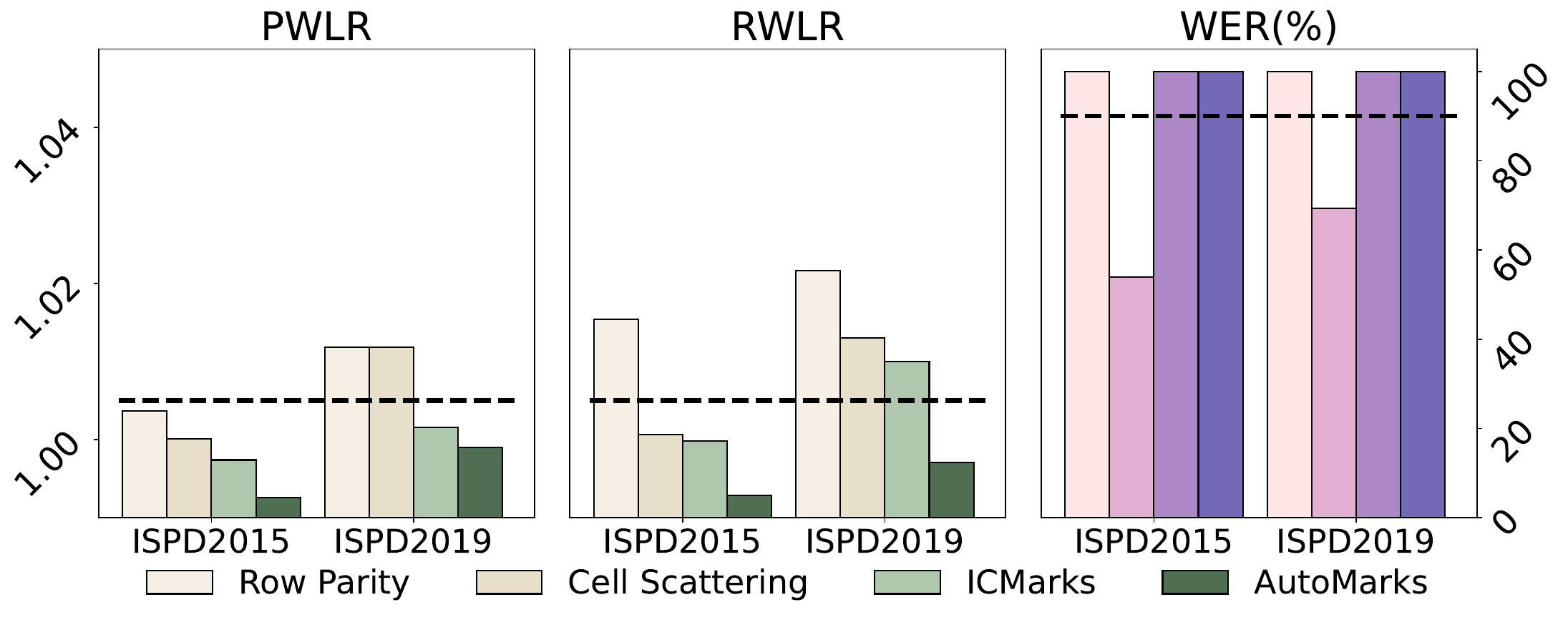}}
    \caption{Watermarking performance under different attacks for wirelength-driven placement on the ISPD'2015~\cite{bustany2015ispd} and ISPD'2019~\cite{liu2019ispd} benchmarks. The black dotted line in the two left subfigures denotes the quality degradation threshold of 1.005, and the black dotted line in the rightmost subfigure denotes the watermark extraction threshold of 90\%.}
    \label{fig:attack_results}
\end{figure*}
\subsection{Additional Visualizations}~\label{append:visual}
In this section, we provide additional visualizations of \sys's label and the training procedure.

\begin{figure}[!ht]
    \begin{minipage}[c]{0.47\columnwidth}
        \includegraphics[width=\textwidth]{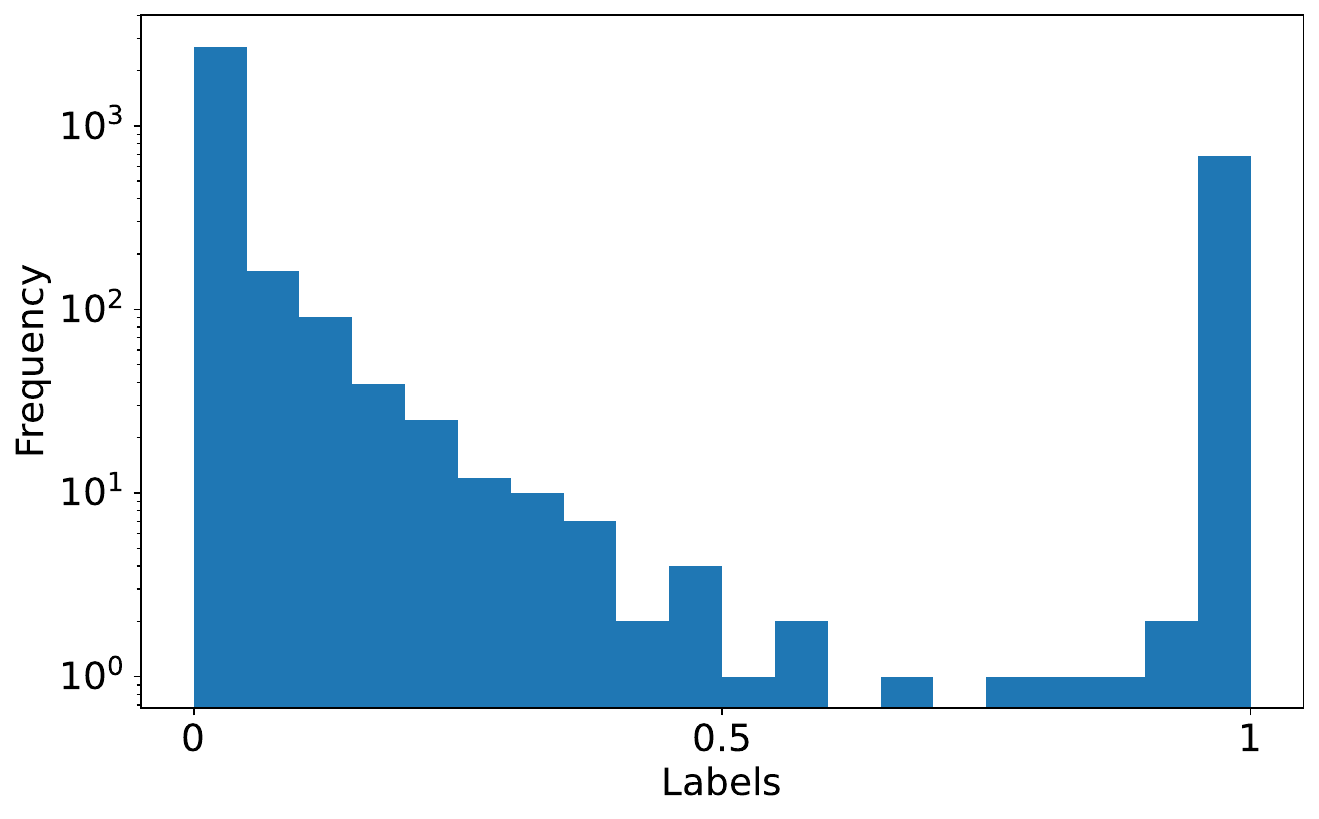}
        \caption{Histogram distribution of \sys's labels. }
        \label{fig:histogram}
    \end{minipage}
    \hfill
    \begin{minipage}[c]{0.47\columnwidth}
        \includegraphics[width=\textwidth]{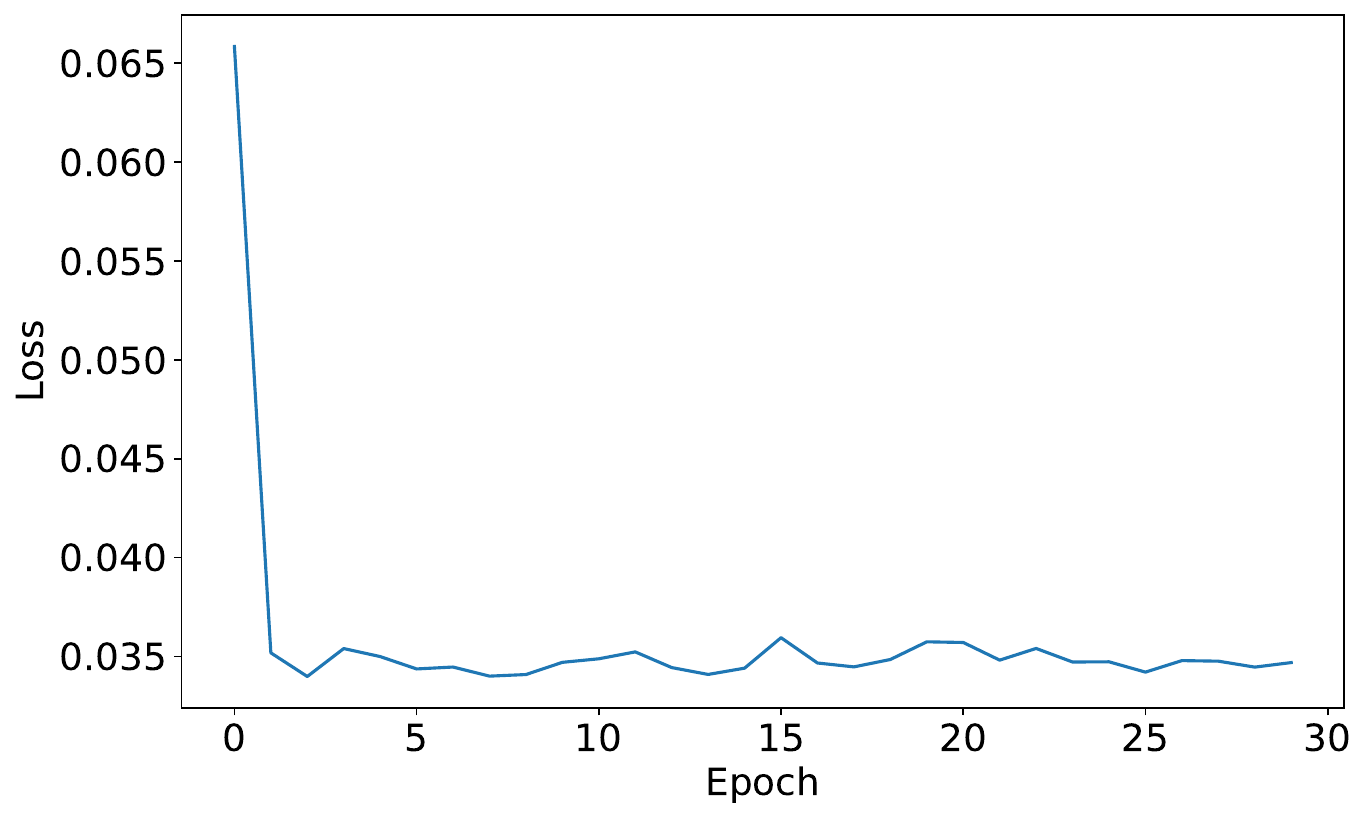}
        \caption{Loss curve during the \sys's GNN training procedure. }
         \label{fig:loss}
    \end{minipage}
    \vspace{-10pt}
\end{figure}

\paragraph{\textbf{Visualization of the labels}}  We  visualize the training dataset \texttt{ispd19test6}'s labels $L$ distribution in Fig~\ref{fig:histogram}. As seen, watermarking on more than half of the nodes yields good watermarking performance. The critical step is to ensure the graph neural network learns the subgraph structure of the good nodes and can thus ensure the watermarking fidelity.  

\paragraph{\textbf{Loss curve during GNN training}}
We show the loss curve in Fig.~\ref{fig:loss}, which demonstrates the graph neural network gradually learns to predict quality degradations of a given subgraph. The loss converges during the GNN training.

\begin{figure}[!ht]
\vspace{-15pt}
    \centering
    \subfloat[test4(WM)]{\label{f:design9} \includegraphics[width=0.2\columnwidth]{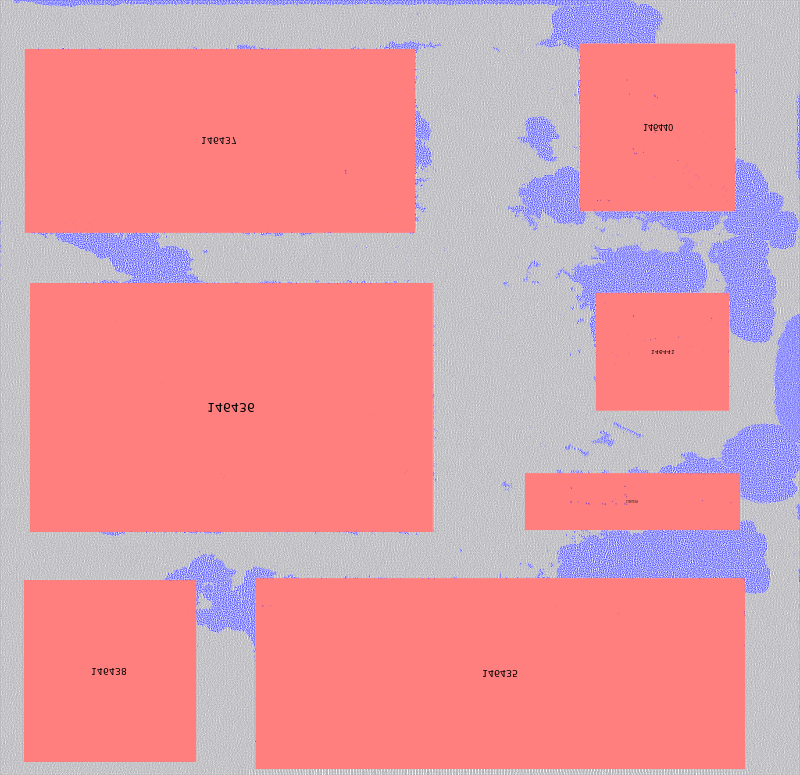}}
    \subfloat[test4(non-WM)]{\label{f:design9} \includegraphics[width=0.2\columnwidth]{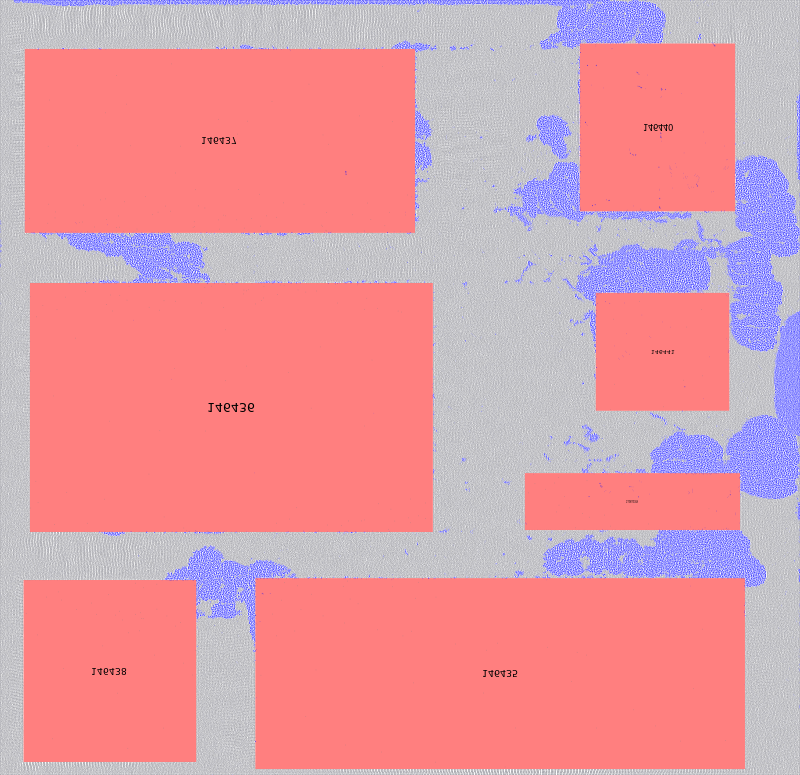}}
    \subfloat[dist\_a(WM)]{\label{f:design9} \includegraphics[width=0.2\columnwidth]{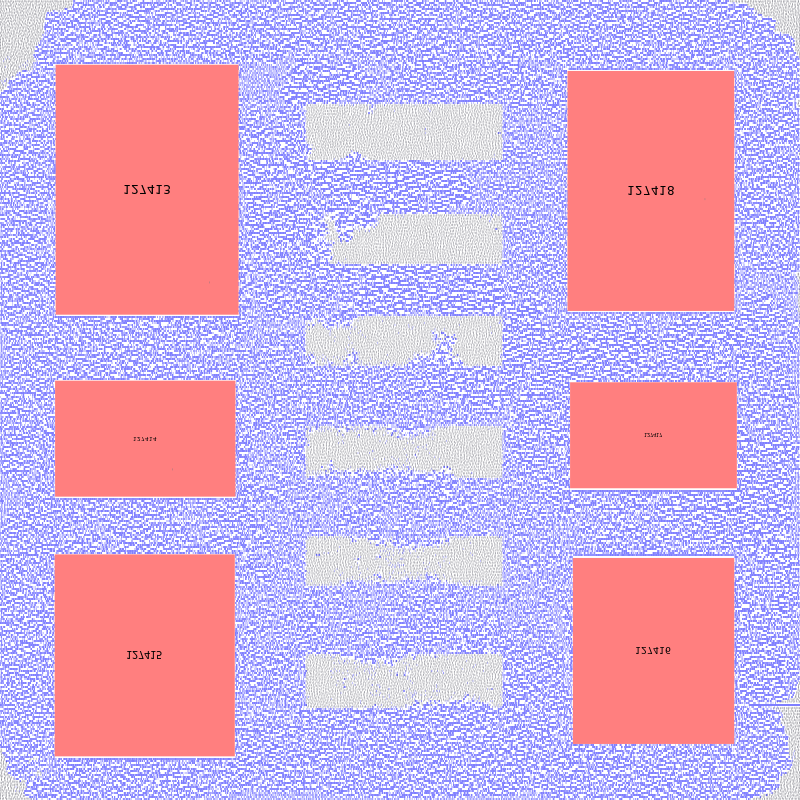}}
    \subfloat[dist\_a(non-WM)]{\label{f:design9} \includegraphics[width=0.2\columnwidth]{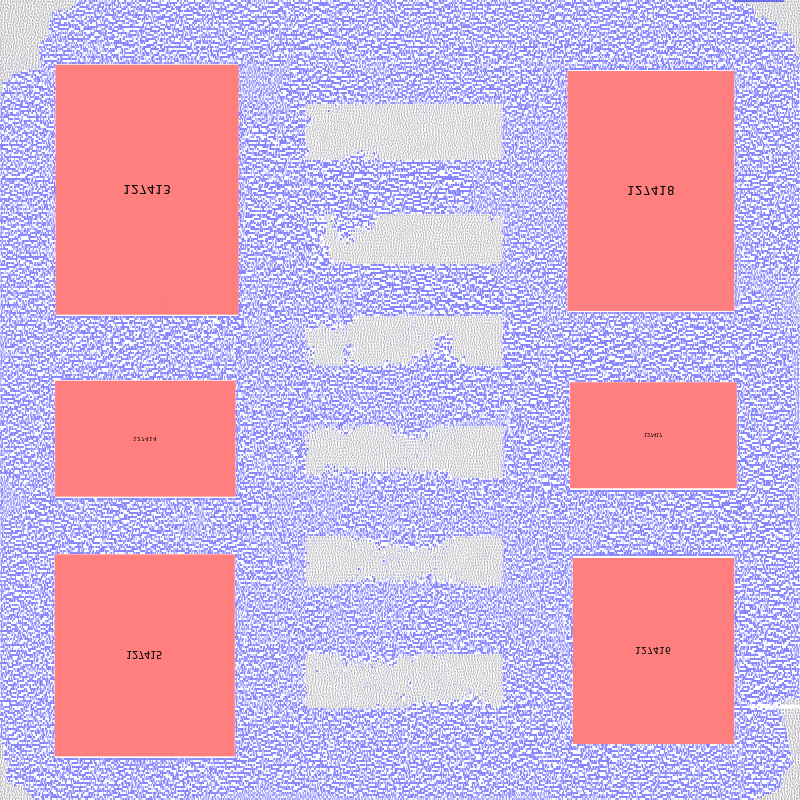}}
  \vspace{-10pt}
    \caption{Watermarked and Non-watermarked examples.}
 \vspace{-10pt}
    \label{fig:example}
\end{figure}

\paragraph{\textbf{ Stealthiness}}~\label{sec:stealthiness} To verify the stealthiness of \sys, we include the watermarked layout and the non-watermarked layout from both ISPD'2015~\cite{bustany2015ispd} and ISPD'2019~\cite{liu2019ispd} benchmarks in Fig.~\ref{fig:example}. As seen, the watermarks are invisible upon inspection and the adversary cannot differentiate between a non-watermarked layout and a layout watermarked by \sys{}.

%
%
 
\subsection{Additional Ablation Studies}~\label{append:ablation}
\paragraph{\textbf{The Effectiveness of learning rates}}
We analyze how different GNN learning rates (LR) would affect \sys's performance in Table~\ref{tab:effectiveness}. Following Sec.~\ref{sec:results}, the \sys{} is trained on \texttt{ispd19test6} design and inference on the rest of the benchmarks. Here, we change the learning rate from 0.01 to 0.001 and 0.1.  When the learning rate is increased from 0.001 to 0.01, the GNN learns to capture the node connections and predict the watermarking fidelity better. However, increasing the learning rate from 0.01 to 0.1 introduced marginal quality changes.
%
%
%
\begin{table}[ht]
\vspace{-10pt}
    \centering
    \begin{tabular}{|c|c|c|}
    \hline
   LR & PWLR & RWLR \\\hline
     0.001 & 1.0004& 1.0009\\
                     0.01  & 0.9981 & 0.9982\\
                             0.1 &0.9990 & 0.9998\\\hline
    \end{tabular} 
    \caption{The effectiveness of different learning rate choices on \sys's performance.}
   \vspace{-30pt}
    \label{tab:effectiveness}
\end{table}
\paragraph{\textbf{The Effectiveness of $\beta$ Threshold}} We analyze how different label threshold $\beta$ choice would impact the watermarking fidelity performance. In Table~\ref{tab:beta}, we change the $\beta$ from 0.01 to 0.005 and 0.2, and show the PWLR and RWLR performance. Other settings follow the default ones in \sys.

As seen, increasing the $\beta$ from 0.005 to 0.01 introduced significant PWLR and RWLR performance improvemnt. It is because more training nodes are normalized into the range of [0,1] that help \sys{} to learn more diverse score predictions. However, increasing the $\beta$ from 0.01 to 0.02 does not introduce significant watermarking fidelity changes. 
\begin{table}[!ht]
\vspace{-10pt}
    \centering
    \begin{tabular}{|c|c|c|}
    \hline
        $\beta$& PWLR & RWLR \\\hline
       0.005 & 1.0078 & 1.0089 \\
       0.01 & 0.9981 & 0.9982\\ 
       0.02 & 1.0013& 0.9990 \\\hline
    \end{tabular} 
    \caption{The effectiveness of different $\beta$ choices on \sys's performance.}\vspace{-30pt}
    \label{tab:beta}
\end{table}
\paragraph{\textbf{The effectiveness of $\gamma$}}  We analyze how different $\gamma$ choices would affect the larger designs' (i.e. cell size larger than 900k) performance in Table~\ref{tab:gamma}.  Other settings follow the default ones in \sys. As seen, adjusting the $\gamma$ from 0.2 to 0 and 0.5 will change the \sys{} performance on the large layout slightly. 
\begin{table}[!h]
\vspace{-10pt}
    \centering
    \begin{tabular}{|c|c|c|}
    \hline
        $\gamma$ & PWLR & RWLR \\\hline
        0 & 0.9974 & 1.0002 \\
        0.2 & 0.9972 & 1.0001\\
       0.5 & 0.9974 &  1.0002\\\hline
    \end{tabular} 
    \caption{The effectiveness of different $\gamma$ choices on \sys's performance.}\vspace{-30pt}
    \label{tab:gamma}
\end{table}

\subsubsection{\textbf{The effectiveness of GNN layers}}  We analyze how different numbers of the GNN layers would affect the watermarking fidelity performance in Table~\ref{tab:layer}. Here, we fix the watermark region size $N=10$ to be the same as Section~\ref{sec:results} and change the number of GNN layers used for training and inference from $5 \rightarrow 10$.

Increasing the layer from 5 to 7 significantly improved the watermarking performance. When the layer number is set to 5, most of the cells are within the watermark region, and the GNN fails to learn the node features on the region boundary, which will be expelled outside the region and play an essential role in quality degradation. Increasing the number of layers results in better watermarking performance. However, increasing the number from 7 to 10 introduced marginal quality improvement and more computational overheads. Therefore, \sys{} employs a graph neural network with 7 layers.
\begin{table}[!ht]
    \centering
    \begin{tabular}{|c|c|c|}
    \hline
        Layer Num. & PWLR & RWLR \\\hline
        5 & 1.1033 &1.1131 \\
       7 & 0.9981 & 0.9982 \\
       10 & 0.9997 & 0.9999\\\hline
    \end{tabular} 
    \caption{The effectiveness of different GNN layers on \sys's performance.}
    \vspace{-30pt}
    \label{tab:layer}
\end{table}

\end{document}